\documentclass[twocolumn,preprintnumbers,amsmath,amssymb]{revtex4-2}
\usepackage{graphicx}
\usepackage[normalem]{ulem}
\usepackage[svgnames]{xcolor}
\usepackage{bm}
\usepackage{multirow}
\usepackage{float}
\usepackage{titlesec}
\usepackage[columnwise]{lineno}
\usepackage{float}
\usepackage{titlesec}
\usepackage{gensymb}
\usepackage{textcomp}

\begin{document}

\title{An electronic band sculpted by oxygen vacancies and indispensable for dilute superconductivity}

\author{Beno\^{\i}t Fauqu\'e$^{1}$}
\author{Cl\'ement Collignon$^{1,2}$}
\thanks{Present address: Department of Physics, Massachusetts Institute of Technology, Cambridge, Massachusetts 02139, USA}
\author{Hyeok Yoon$^{3,4}$}
\thanks{Present address: Department of Physics, University of Maryland, College Park, Maryland 20742, USA}
\author{Ravi$^{2}$}
\author{Xiao Lin $^{2}$}\thanks{Present address: School of Science, Westlake University, 18 Shilongshan Road, 310024 Hangzhou, China}
\author{Igor I. Mazin$^{5}$}
\author{Harold Y. Hwang$^{3,4}$}
\author{Kamran Behnia$^{2}$}

\affiliation{
(1) JEIP  (USR 3573 CNRS), Coll\`ege de France,  75005 Paris, France\\
(2) Laboratoire de Physique et d'Etude de Mat\'{e}riaux (CNRS)\\ ESPCI Paris, Université PSL, 75005 Paris, France\\
(3) Stanford Institute for Materials and Energy Sciences, SLAC National Accelerator Laboratory, Menlo Park, Stanford, California 94025, USA\\
(4) Department of Applied Physics, Stanford University, Stanford, California 94305, USA\\
(5) Department of Physics and Astronomy and Quantum Science and Engineering Center, George Mason University, Fairfax, Virginia 22030, USA
}

\date{\today}
\begin{abstract}
Dilute superconductivity survives in bulk strontium titanate when the Fermi temperature falls well below the Debye temperature. Here, we show that the onset of the superconducting dome is dopant-dependent. When mobile electrons are introduced by removing oxygen atoms, the superconducting transition survives down to $2 \times 10^{17}$ cm$^{-3}$, but when they are brought by substituting Ti with Nb, the threshold density for superconductivity is an order of magnitude higher. Our study of quantum oscillations reveals a difference in the band dispersion between the dilute metals made by these doping routes and our band calculations demonstrate that the rigid band approximation does not hold when mobile electrons are introduced by oxygen vacancies. We identify the band sculpted by these vacancies as the exclusive locus of superconducting instability in the ultra-dilute limit.
\end{abstract}
\maketitle
\section{Introduction}
Discovered half a century ago \cite{Schooley1964}, superconductivity in strontium titanate is attracting renewed attention \cite{CollignonRev2019,GASTIASORO}. It lies beyond the  Migdal-Eliashberg \cite{eliashberg1960} framework, since the Fermi temperature of electrons falls below the Debye temperature of phonons \cite{Lin2013,Yoon2021}. The dilute metal going through this superconducting instability has non-trivial transport properties \cite{Collignon2020,Kumar2021,Nazaryan2021,Collignon2021}. The insulating parent, identified as a quantum paraelectric \cite{Muller1979}, has a huge permittivity, displays an unusual low-temperature thermal transport \cite{Martelli2018} including a thermal Hall component \cite{Li2020}. Soft transverse optical phonons, hybridizing with transverse acoustic phonons at low temperatures \cite{Delaire2020, Fauque2022}, are suspected to play a major role.

The proximity to a ferroelectric instability is often invoked in explaining  superconductivity \cite{Rowley2014,Edge2015, Enderlein2020}. Several experimental studies have documented a constructive interplay between superconductivity and ferroelectricity \cite{Rischau2017,Tomioka2019,Ahadi2019,Rischau2022} and recent theoretical accounts of Cooper pairing invoke the exchange of two \cite{Marel2019,Kiselov2021,Volkov2022} or one \cite{Gastiasoro2022,Yu2022} soft phonons between pairing electrons. A two-phonon exchange scenario, first invoked decades ago \cite{Ngai1974}, leads to instantaneous electron-electron attraction and can explain \cite{Kiselov2021} the experimentally-observed persistence of superconductivity at  densities below 10$^{18}$ cm$^{-3}$ \cite{Lin2013,Lin2014}.   

Here, we present a study of superconducting transition and quantum oscillations in oxygen-reduced and Nb-substituted strontium titanate in the dilute limit. We find that when $n < 5 \times 10^{18}$ cm$^{-3}$, superconductivity is present in SrTiO$_{3-\delta}$, but absent in SrTi$_{1-x}$Nb$_x$O$_3$, either single-crystalline or custom-made thin films. The dilute metals generated by alternative doping routes are also different. In particular, the Lifshitz transition in  SrTi$_{1-x}$Nb$_x$O$_3$ occurs at a lower density.  We present numerical evidence that the virtual crystal approximation used in the LDA band calculations \cite{marel2011} holds for Nb doping, but not for oxygen deficiency. We conclude that while both doping routes eventually lead to a superconducting dome, dilute superconductivity (arising in a metal where e-e distance is of the order 10 nm) occurs only when the occupied band is sculpted by oxygen vacancies. A larger coupling between the electrons of the vacancy-sculpted band and soft TO phonons, or a larger electronic density of states, or a combination of both, may be the origin.

\begin{figure}[ht]
\begin{center}
\includegraphics[width=8
cm]{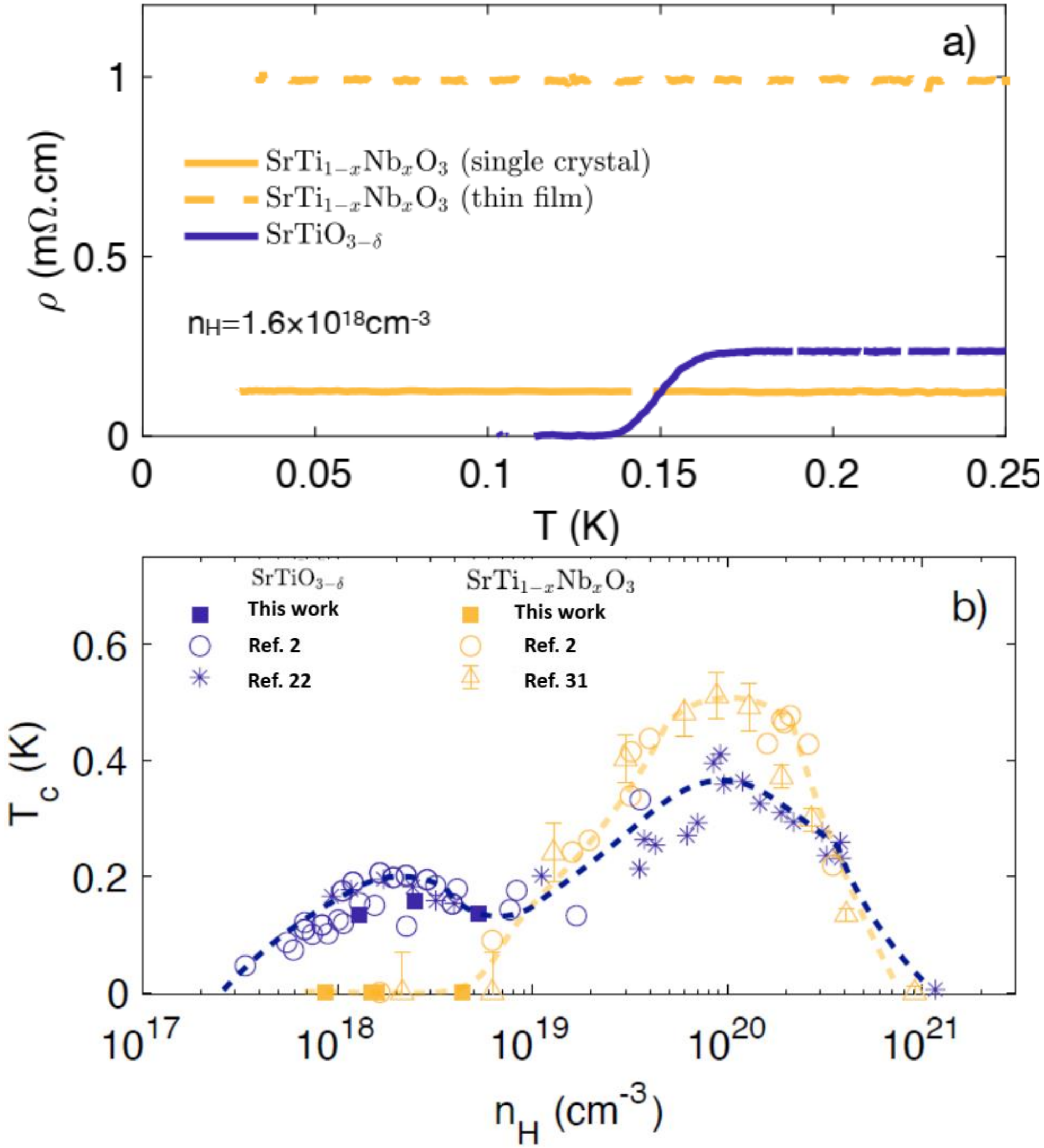}
\caption{\textbf{Different thresholds for emergence of superconductivity: } 
a) Temperature dependence of electrical resistivity ($\rho_{xx}$) in SrTi$_{1-x}$Nb$_x$O$_{3}$ and SrTiO$_{3-\delta}$ (orange) sample with similar carrier density n$_H\simeq$1.6 $\times 10^{18}$cm$^{-3}$. Neither the thin film nor the single crystalline SrTi$_{1-x}$Nb$_x$O$_{3}$ shows a superconducting transition at this carrier density.  b) Doping evolution of the superconducting critical temperature (T$_c$). Solid symbols represent samples studied in the present work. Open symbols represent what was reported previously \cite{CollignonRev2019,Rischau2022,Tomioka2022}. }
\label{FigDomes}
\end{center}
\end{figure}

%A recent DFT+DMFT calculation  by Souto-Casares, Spaldin and Ederer \cite{souto2021} found that, as suspected previously \cite{Hou2010}, the consequences of substituting a Ti atom by Nb is quite different to removing oxygen. In the first case, the lower conduction band of the stoichiometric solid, which is rooted in Ti orbitals, is filled by the single electron introduced by the dopant. In contrast, the fate of two electrons brought by an oxygen vacancy is more eventful. Electron-electron repulsion is sizable both on the vacancy site and on the Ti site and two alternatives compete \cite{souto2021}. In the first,  a localized vacancy state is doubly occupied. In the second, the vacancy is occupied by a single electron while the other dopes the conduction band. In this latter case, the singly occupied in-gap state is the lower Hubbard band on the vacancy site and the mobile electron is in the higher Hubbard band. The latter, which has a minimum at the $\Gamma-$ point like the lower conduction band of stoichiometric SrTiO$_{3}$ yet a different dispersion, becomes the locus of metallicity. Our study of comparative fermiology of these two dilute metals in broad agreement with this site-selective `Mottness' picture \cite{souto2021}. Note, however, that the range of substitution scanned by our experiment is beneath $10^{-3}$, far below what can be theoretically probed. Moreover, these calculations were restricted to the cubic phase, neglecting the lifting of the threefold degeneracy of the electronic bands by spin-orbit coupling and tetragonal distortion \cite{marel2011}.

SrTiO$_3$ can be n-doped by a variety of atomic substitutions. The focus of the present study is to compare substitution of  tetravalent Ti$^{4+}$ by pentavalent  Nb$^{5+}$ with creation of oxygen vacancies. While both lead to n-doping, there is no {\it a priori} reason to believe that they affect the electronic structure in the same way \cite{Hou2010}. 
%A first impression can be drawn starting from infinitesimally small concentration of just one doped electron. In that case, correlation effects can be neglected (an electron does not correlate with itself). In case of Nb the electron can be bound to the Nb$^{+3}$ ion, gaining potential energy, or it can be fully itinerant, gaining kinetic energy. Since this is a solely one-electron effect, so it can be, if it exists, captured by density functional (DFT) calculations. On the contrary, O vacancy means that the two neighboring Nb ions are left with one extra electron each. What can happen with these electrons? First, they can get collectivized over all Ti sites. Second, one or both can occupy the vacancy (this scenario was discussed in Ref. \cite{souto2021}). However, barring correlations among the doped electrons, there is little incentive for them to get localized. Coulomb repulsion works against localizing them both within the vacancy site, as also found in Ref. \cite{souto2021}. Moreover, since the vacancy is, in the first approximation, neutral, and the Ti ions positively charges, an individual electron would never leave the Ti network for a vacancy.
In presence of an O vacancy, the two neighboring Ti ions are left with one extra electron each. What can happen with these electrons? They can either get collectively extended over all Ti sites or both occupy the vacancy. %However, barring correlations among the doped electrons,  there is little incentive for them to get localized. 
In a recent DFT+DMFT calculation, Souto-Casares, Spaldin and Ederer \cite{souto2021} found that if the Hubbard repulsion on Ti $and$ the concentration are both large, one or even two electrons will prefer to avoid the Hubbard repulsion on Ti and ``hide'' in the vacancy. However, the effective concentration of vacancies in that work \cite{souto2021} was one per four formula units, which is several orders of magnitude higher than the relevant experimental range. At very small concentrations, Hubbard correlation between the doped electrons, included in the DMFT method and not in the DFT, are  much less important than geometric changes such as lattice relaxation. Keeping this in mind, we have performed full structural relaxation in the density functional theory (DFT), using an $8\times8\times8$ supercell of 64 f.u., and either replacing one Ti with Nb  or removing one O. As discussed below, we found that the resulting electronic structure near the bottom of the conduction band is dramatically different in the two cases.

\begin{figure}[ht]
\begin{center}
\includegraphics[angle=0,width=8cm]{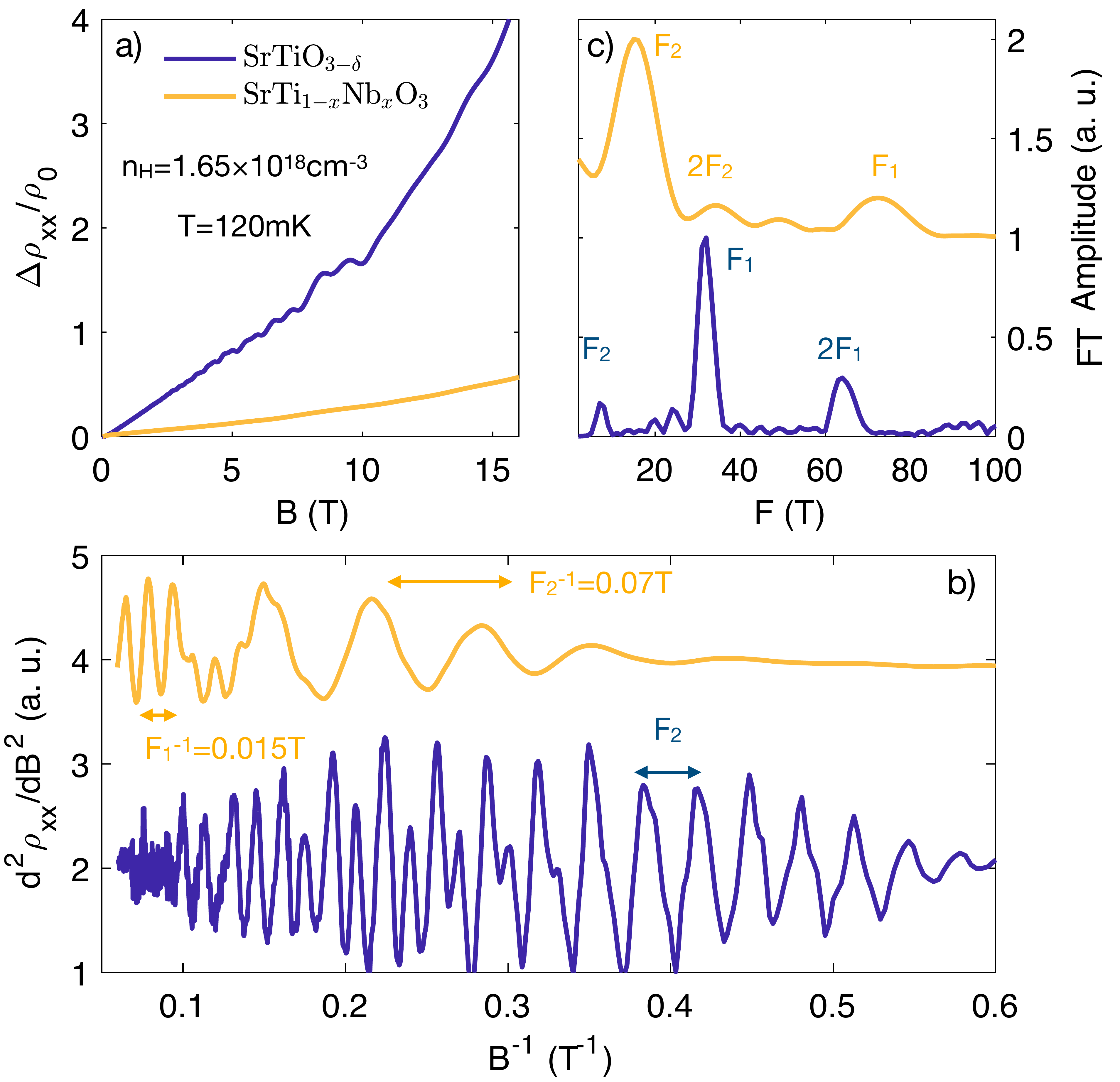}
\caption{\textbf{Distinct Fermi surfaces:} a) Magnetoresistance ($\frac{\Delta \rho_{xx}}{\rho_0}$) vs B in SrTi$_{1-x}$Nb$_x$O$_{3}$ (orange) and SrTiO$_{3-\delta}$ (blue) with the same Hall carrier density n$_H$=1.65e18cm$^{-3}$ at T = 120 mK. b) Traces of the quantum oscillations in $\frac{\partial^2\rho_{xx}}{\partial B^2}$ as function of B$^{-1}$ for both samples. c) Amplitude of the Fourier transform (FT) vs frequency (F) deduced from b). The amplitude was normalized by the peak value and shift for clarity. In both samples, two frequencies (and their harmonics) can be identified, respectively labelled F$_1$ and F$_2$.}
\label{FigQO1}
\end{center}
\end{figure}

\begin{figure}[ht]
\begin{center}
\includegraphics[angle=0,width=8cm]{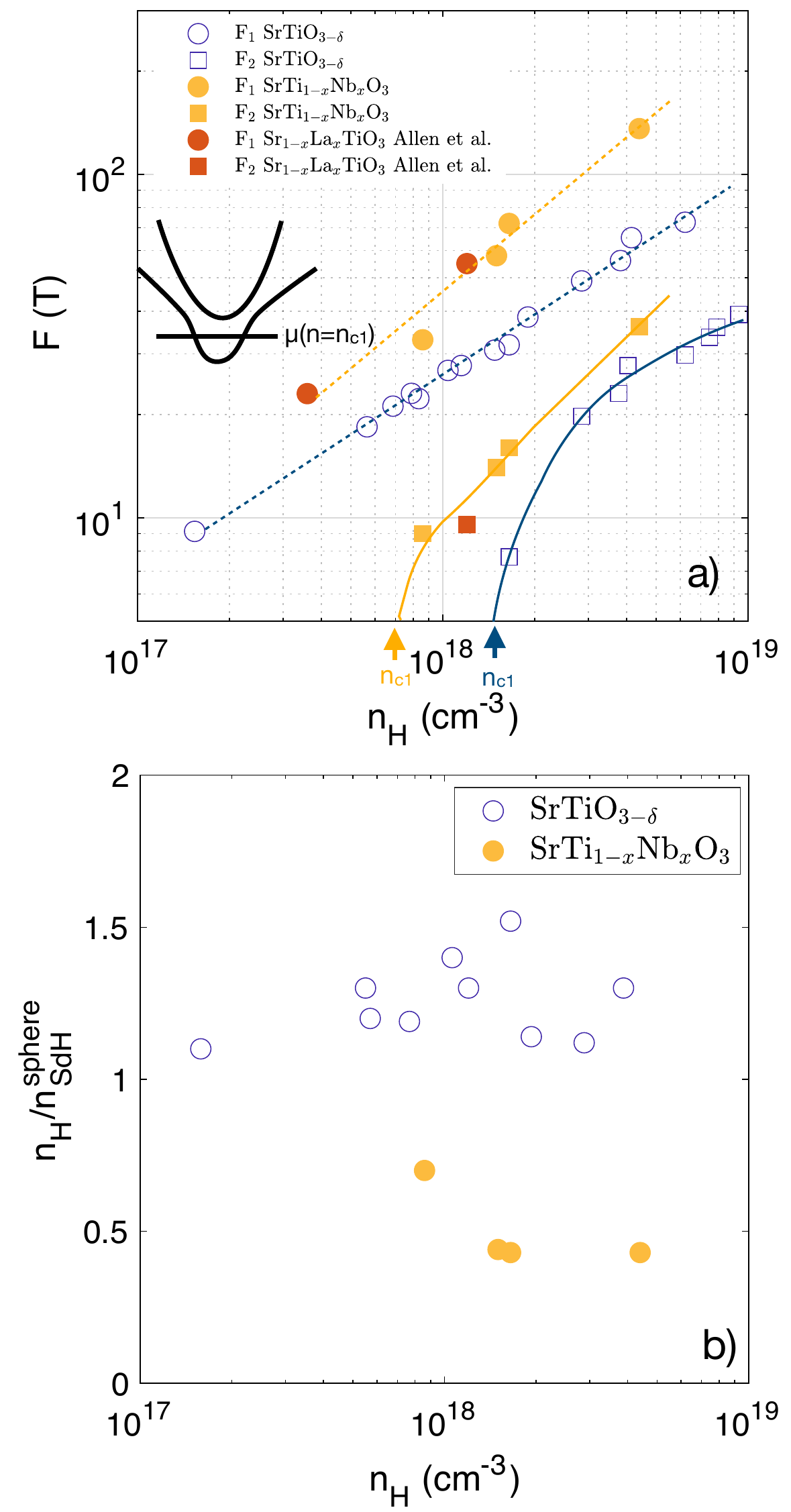}
\caption{\textbf{Doping evolution of the Fermi surface of SrTi$_{1-x}$Nb$_x$O$_{3}$ and SrTiO$_{3-\delta}$ :} a) Frequencies of the quantum oscillations  of the two sub-bands F$_1$ (circles) and F$_2$ (squares) \textit{vs.} the Hall carrier density (n$_H$) in SrTi$_{1-x}$Nb$_x$O$_{3}$, in SrTiO$_{3-\delta}$ \cite{Lin2014} and in Sr$_{1-x}$La$_x$TiO$_{3}$ \cite{Allen2013}. The Lifshitz transition occurs at different threshold densities. b) Ratio of the carrier density extracted from the Hall coefficient to the carrier density deduced from quantum oscillations frequency assuming two spherical pockets. A difference persists in all samples. This implies distinct geometries for the Fermi surface in the two cases.}
\label{FigFreqvsn}
\end{center}
\end{figure}

\begin{figure}[ht]
\begin{center}
\includegraphics[angle=0,width=8cm]{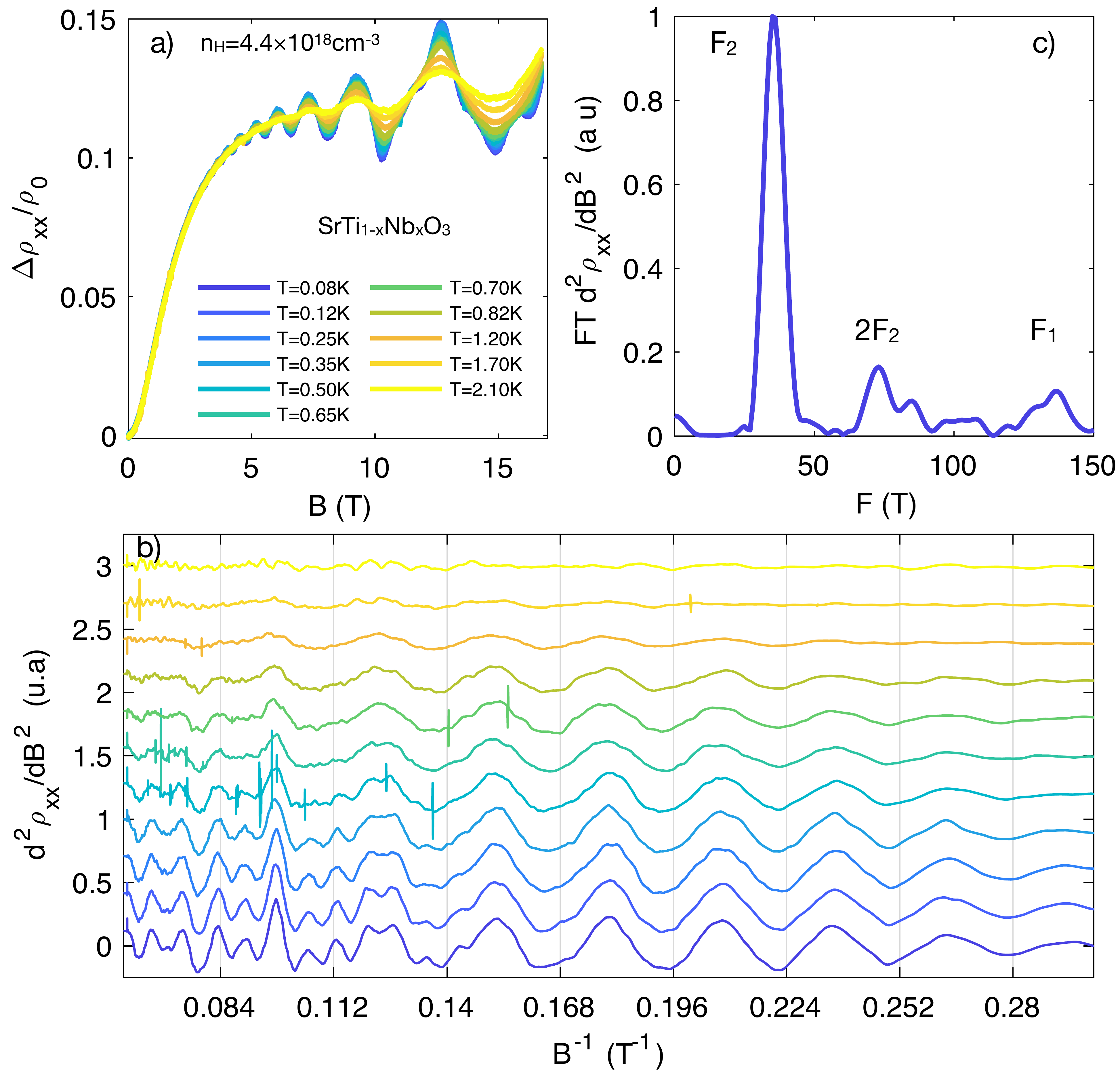}
\caption{\textbf{Quantum oscillations in SrTi$_{1-x}$Nb$_x$O$_{3}$  with n$_H=4.4\times10^{18}$cm$^{-3}$ :}  a) $\frac{\Delta \rho_{xx}}{\rho_0}$ vs B at different temperatures from 80 mK to 2.1 K. b)  $\frac{\partial^2\rho_{xx}}{\partial B^2}$ as a function of B$^{-1}$ displaying quantum oscillations. c) FT of the data at T = 80 mK. }
\label{FigQO2}
\end{center}
\end{figure}
%Inset : Temperature dependence of the amplitude of the oscillation at B = 10.5 T. The red line corresponds to a fit using the Lifshitz-Kosevich formula with $m^{\star}$=1.4$m_0$.
\section{Results}
 \subsection{Dopant dependent superconductivity} Fig.\ref{FigDomes}a shows the low-temperature electrical resistivity ($\rho_{xx}$) of three samples with the same Hall carrier density. One can see that SrTiO$_{3-\delta}$ becomes a superconductor, but not SrTi$_{1-x}$Nb$_{x}$TiO$_3$  (at least not down to 30mK). Note that, the oxygen deficient sample has a residual resistivity lower than the thin film and larger than the crystalline  SrTi$_{1-x}$Nb$_{x}$TiO$_3$. This rules out disorder as the driver of this dichotomy. Studying four different Nb-doped SrTiO$_3$ thin films \cite{Kozuka2010} with a thickness of $\approx$ 900 nm and nominal Nb concentrations of 0.02 and 0.04 atomic percent, we found that none was superconducting (See the supplement for more details about samples and data \cite{SM}). 

Fig.\ref{FigDomes}.b draws the superconducting phase diagram of SrTiO$_{3-\delta}$ and in SrTi$_{1-x}$Nb$_{x}$TiO$_3$ according to the available data reported by different groups \cite{CollignonRev2019, Rischau2022, Tomioka2022}. The critical temperature does not evolve identically. The peak in T$_c$ of the principal dome is higher in Nb-doped strontium titanate, while a lower superconducting dome is present for SrTiO$_{3-\delta}$ and absent in SrTi$_{1-x}$Nb$_{x}$O$_3$. 

Evidence for superconductivity in this density range is limited to the detection of zero resistivity. The  critical temperature of bulk probes of superconductivity, such as specific heat \cite{Lin2014b}, thermal conductivity  \cite{Lin2014b} and diamagnetism  \cite{Collignon2017} is consistently lower than the temperature at which resistivity drops to zero. Given the strong variation of critical temperature with pressure or strain \cite{Enderlein2020}, this points to the survival of filamentary superconductivity with a higher critical temperature at specific locations like dislocations or domain boundaries. Nevertheless, oxygen-reduced samples display bulk superconductivity for a carrier density as low as 4.5$\times 10^{18}$cm$^{-3}$ \cite{Rischau2017,Rischau2022} and the Nb-substituted ones do not show any transition (resistive or else) at this concentration. Therefore, there is a genuine difference of the ground states, irrespective of the origin of filamentary superconductivity.

 \subsection{Dopant dependent metallicity}
Let us now compare the normal state of these two dilute metals. Structurally, the distribution of oxygen vacancies is known to be less homogeneous than the distribution of Nb atoms \cite{Szot2002}. However, in both cases, electrons belong to a single Fermi sea (and not to a plurality of puddles). Since the Bohr radius is as long as several hundreds of nanometer \cite{Behnia2015}, disorder is averaged over a volume containing many  dopants. In both oxygen-reduced and Nd-substituted strontium titanate, the carrier mean-free-path is much longer than the interdopant distance and quantum oscillations are visible at moderate ($\approx 2$ T) magnetic fields.

We carried out a careful examination of the Fermi surface by looking at the Shubnikov-de Haas effect in the two cases.  Magnetoresistance  ($\frac{\Delta \rho_{xx}}{\rho_0}$) at T = 120 mK and for B$\parallel$ [001] (See Fig.\ref{FigQO1}a)) reveals a difference. Quantum oscillations are visible on top of a monotonic background in both. As seen in Fig.\ref{FigQO1}b, however, despite the quasi-equality of their Hall carrier density, the second derivative of resistivity $\frac{\partial^2\rho_{xx}}{\partial B^2}$ shows different patterns. Two main frequencies, labelled F$_1$ and F$_2$ can be resolved, in qualitative agreement with previous studies of quantum oscillations \cite{Uwe_1985,Allen2013,Lin2013}. However, F$_1$ and F$_2$ are not identical in the two cases, implying a difference in the structure of the Fermi surface between superconducting SrTiO$_{3-\delta}$ and non-superconducting SrTi$_{1-x}$Nb$_{x}$TiO$_3$. %Note that the onset magnetic field for quantum oscillations is lower in the oxygen-reduced sample indicating a longer Dingle time, despite a shorter transport scattering time.

By studying three other Nb-doped thin-film samples with n$_H$ ranging from 8.6$\times 10^{17}$ cm$^{-3}$ to  4.4$\times 10^{18}$cm$^{-3}$ (see \cite{SM} for sample details), we found that the evolution of F$_1$ and F$_2$ in SrTi$_{1-x}$Nb$_x$O$_{3}$ and in SrTiO$_{3-x}$ \cite{Lin2014} are not identical (See Fig. \ref{FigFreqvsn}.a). At a given Hall carrier density, both frequencies  are larger in SrTi$_{1-x}$Nb$_x$O$_{3}$ compared to SrTiO$_{3-\delta}$ (See the supplement \cite{SM} for more details on the analysis).  
 
In the absence of an angle-dependent study, possible multiplicity of structural domains below 105 K is a source of complication. In multi-domain samples the orientation of magnetic field can be perpendicular or parallel to the c-axis according to the orientation of domains. However, a simple procedure reveals a difference in Fermi surface geometry. Assuming that the two sub-bands are isotropic, the carrier density can be deduced from the frequencies of quantum oscillations: n$^{sphere}_{SdH}$=$\frac{1}{3\pi^2}((\frac{2eF_1}{\hbar})^{3/2}+(\frac{2eF_2}{\hbar})^{3/2}$) and then compared to the Hall carrier density, $n_H$. Their ratio,  $\frac{n_H}{n^{sphere}_{SdH}}$, is a measure of the deviation of the pockets from perfect spheres. Fig. \ref{FigFreqvsn}.b compares the two cases at different doping levels. The ratio is $\approx 1.2$ in SrTiO$_{3-\delta}$ and $\approx 0.5$ in SrTi$_{1-x}$Nb$_x$O$_{3}$. Thus, the Fermi surface of Nb-doped samples is less spherical than the Fermi surface of SrTi$_{1-x}$Nb$_x$O$_{3}$. Moreover, as seen in Fig. \ref{FigFreqvsn}.a, the threshold for filling the second band, i.e. the onset of the Lifshitz transition is also lower in the Nb-doped case. 

We note that F$_1$ and F$_2$ of our Nb-doped samples are in good agreement with what has been found in La-doped samples \cite{Allen2013} (See Fig. \ref{FigFreqvsn}a)). This suggests that the Fermi surface in La-doped and in Nb-doped strontium titanate are similar and both differ from the one belonging to oxygen-deficient strontium titanate. Interestingly, there is no report of a superconducting transition in single crystals \cite{Tomioka2019} or thin films of Sr$_{1-x}$La$_x$TiO$_{3}$ \cite{Ahadi2019} when the carrier density is as low as ($\approx 10^{18}$ cm$^{-3}$).

By studying the temperature dependence of the quantum oscillations and using the Lifshitz-Kosevitch formalism, we extracted the cyclotron mass of electrons. Fig. \ref{FigQO2} shows the quantum oscillations and their temperature dependence in the sample with highest carrier concentration (n$_H=4.4 \times 10^{18}$cm$^{-3}$). The results of the analysis are shown in the upper panel of Fig. \ref{FigQO2}c). The effective  mass for the lower band is m$^{\star}=1.4\pm0.1 $m$_0$, comparable to the value found in previous studies of Nb-doped \cite{Uwe_1985} (and La-doped \cite{Allen2013}) strontium titanate. The effective mass of the lower band in the oxygen-reduced case at this carrier density is m$^{\star}=1.8$ m$_0$ \cite{Lin2013} \footnote{The lower band is non-parabolic \cite{marel2011}, and therefore, the effective mass is expected to increase with increasing carrier density. Experimentally, m$^{\star}$ in the oxygen-doped samples becomes $\approx$ 3.5 m$_0$ at n$_H= 3 \times 10^{19}$cm$^{-3}$ \cite{Lin2014}}.

\subsection{Band calculations }
Numerical emulation of a regime as dilute as the one explored experimentally requires unrealistically large supercells. Therefore, a quantitative and microscopic explanation of the observed difference is challenging. However, considerable insight can be gained from DFT calculations. To this end, we used the  projected augmented wave pseudo-potential code VASP \cite{vasp} to fully optimize the crystal structure, including the lattice parameters, in the $8\times8\times8$ supercell of 64 formula units.

\begin{figure}[ht]
\begin{center}
\includegraphics[angle=0,width=8.5cm]{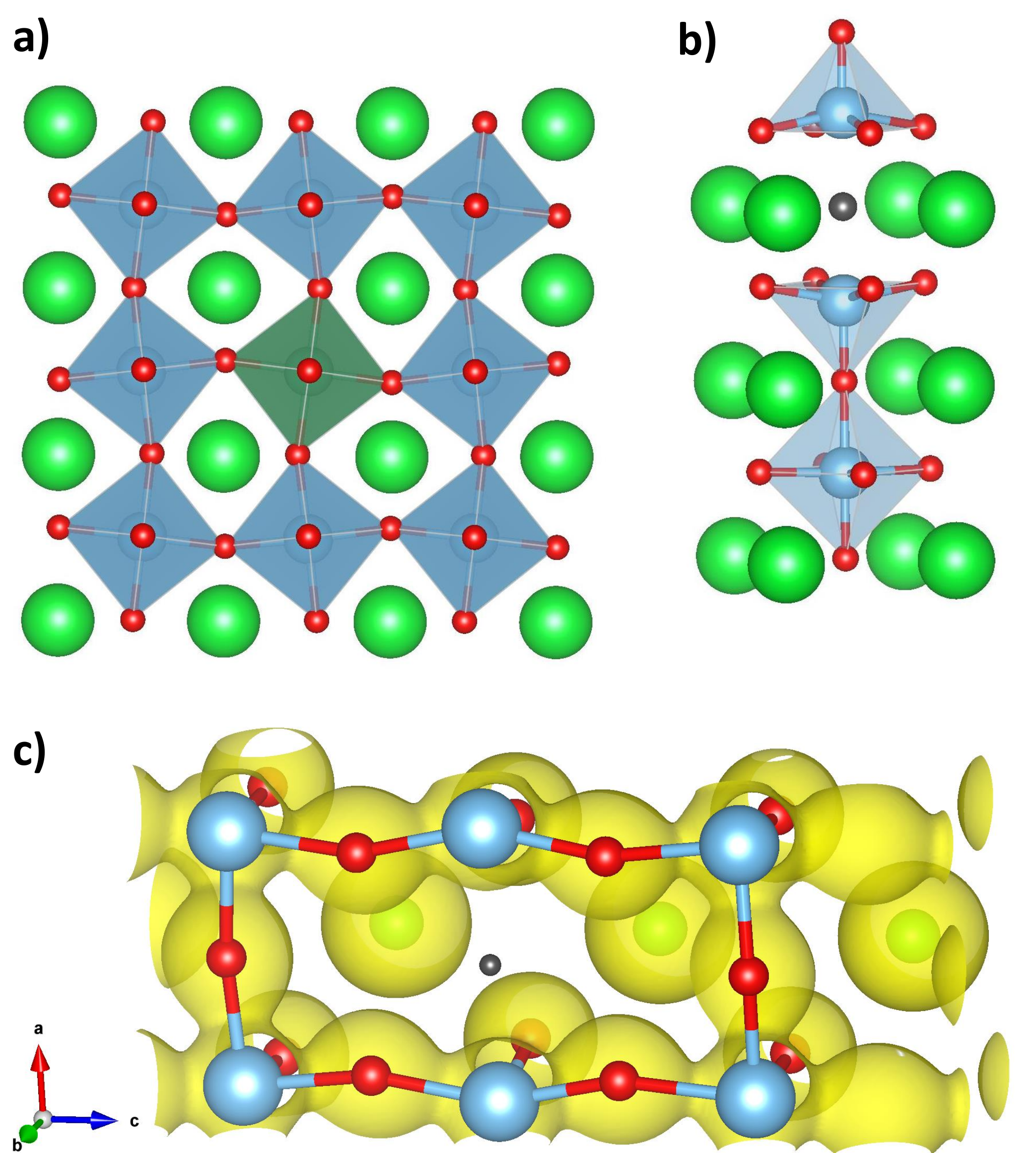}
\caption{\textbf{Effect of impurities on the crystal structure and electron density:} a) Crystal structure with a Nb impurity (green octahedron) b) Same for an O vacancy (black ball) preserving the tetragonal symmetry. c) The electron charge distribution around the vacancy; here the red balls are O, the blue-grey ones Ti and Sr are green. The black ball indicates the approximate location of the vacancy. The (yellow) isosurfaces are drawn at the level of electron density of 0.05 $e/$\AA$^3$. } 
\label{structure}
\end{center}
\end{figure}

In the ground state, at $T=0$, SrTiO$_3$ forms a tetragonal I4/mcm structure, characterized with alternating rotations of the TiO$_6$ octahedra around the $c$ axis. This is a strong effect well captured by the standard DFT calculations (Fig. \ref{structure}). Proper stacking of the rotations along $c$, however, is quite difficult to obtain, and requires sophisticated setups. Our calculations, using standard PBE pseudoptentials, and up to 3x3x3 mesh in the Brillouin zone, reproduce the rotations well, but form a different stacking, corresponding to a P4/mbm group. The difference is not relevant for our analysis.

Since the optimized structure is tetragonal, there are two crystallographically nonequivalent sites, one preserving the tetragonal symmetry (we shall call it v-T) and the other lowering it to an orthorhombic one (v-O). In our calculations, the former corresponds to the space group P4/m (incidentally, removal of Nb also results in the same symmetry), and the latter to Pmm2. At the level of our accuracy we could not determine with certainty which vacancy position is energetically preferable (see the supplement \cite{SM} for details). 

However, we were able to make one structural observation: both Nb and v-T case can be optimized with a restriction forbidding octahedra rotations (v-O cannot). In the undoped compound these rotations decrease the total energy by 3.4 meV per formula. This energy difference increases, in our supercell, to 4.7 meV for Nb substitution. In contrast, it decreases to 0.9 meV for O vacancy. As discussed in the next section, this trend is in agreement with previous experimental findings.

\begin{figure*}[ht]
\begin{center}
\includegraphics[angle=0,width=18cm]{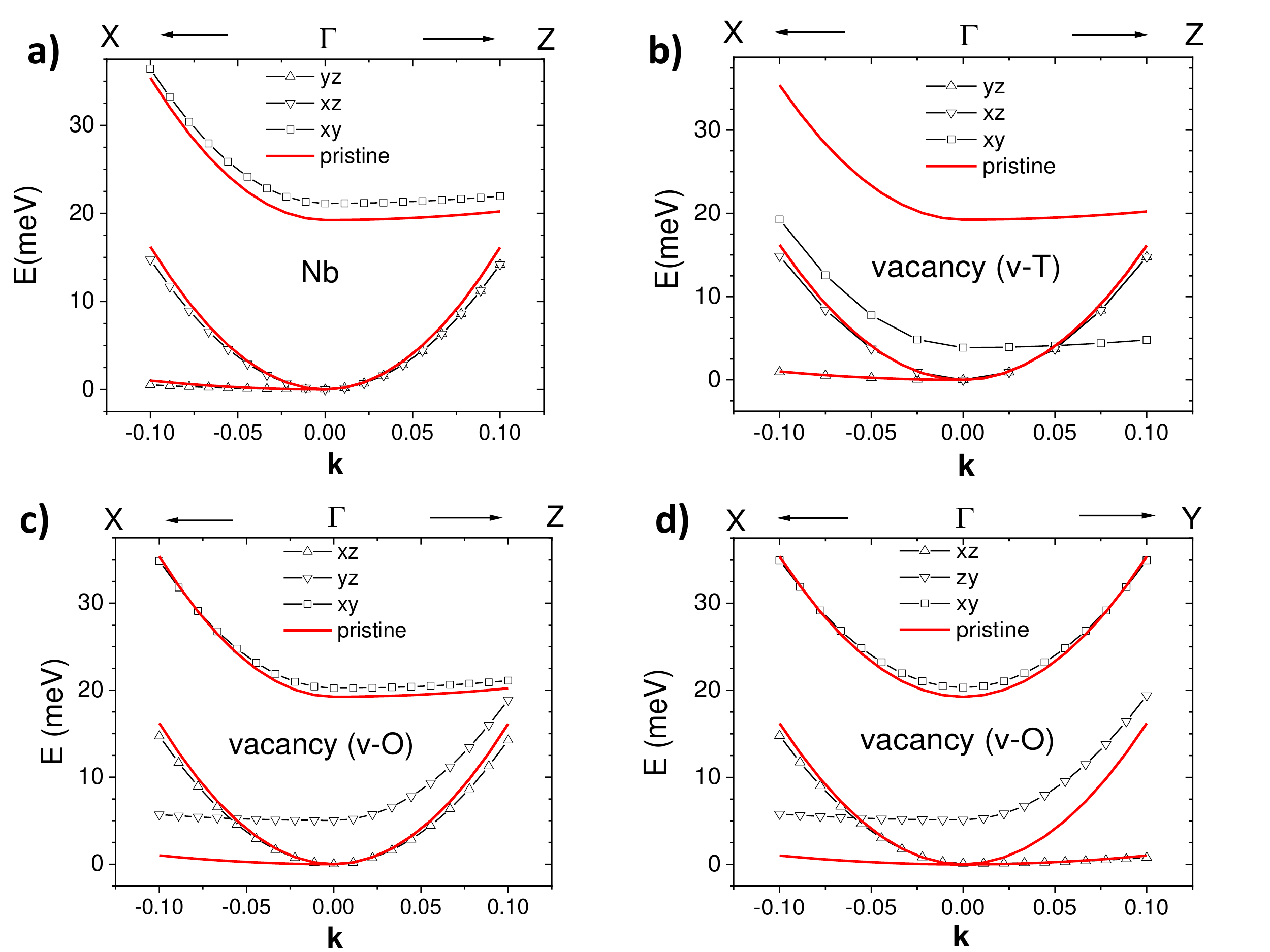}
\caption{\textbf{Non-relativistic band states in
 Sr$_{64}$Ti$_{63}$NbO$_{192}$ (a) and  in Sr$_{64}$Ti$_{64}$O$_{191}$ (b,c,d):} In all cases, the structure was fully optimized within DFT. Only in the case of Nb substitution, the dispersion is virtually identical to the pristine case. In (b), the oxygen vacancy (v-T) keeps the tetragonal symmetry of the lattice. In (c) and (d), the oxygen vacancy (v-O) makes the system orthorhombic and as seen in (d) the dispersion along x- and y-axes are no more identical. The horizontal axis refers to distances in the momentum space along different orientations expressed in reciprocal lattice units. } 
 
\label{Figtheo}
\end{center}
\end{figure*}

The difference in the electronic structure (Fig. \ref{Figtheo}) is even more spectacular. Nb doping has little effect on the band dispersions, but the O vacancies dramatically modifies them. In case of v-T, the main effect is the relative shift between the two lower ($xz$ and $yz$) bands and the upper ($xy$) band.  This can be understood by analyzing the change of the local crystal field. As Fig. \ref{structure} shows, Nb introduces no visible additional distortion, while the vacancy alters the local environment of the two neighboring Ti from octahedral to pyramidal, and leads to a visible relaxations of the eight closest oxygen atoms. 
%Already on the structural level we observe interesting differences between the two cases; the energy gain  between the tetragonal and quasicubic settings is 3.4 meV/formula in the undoped compound, 4.7 meV for Nb, and 0.9 meV for an O vacancy. Assuming that the transition temperature, $T_s$, is proportional to this energy, in a first approximations, we see that an O should lower Tc by 75\% and one Nb increase it by 40\%, \textcolor{red}{qualitatively consistent with the experiment\cite{Tao2016}.}

%The difference becomes even more spectacular when we look at the electronic structure (Fig. \ref{Figtheo}. We see that Nb doping changes the band dispersions rather little, indicating that the rigid-band approximation works well (indeed it was shown previously [Kamran?] that rigid DFT bands describe the Lifshits transitions in Nb-doped STO extremely well). 

%Interestingly, the O vacancy case is dramatically different, the main effect here being the relative shift of the two lower ($xz$ and $yz$) bands and the upper $xy$ band. This can be understood analyzing the change of the local crystal field. As Fig. \ref{structure} shows, Nb introduces no visible additional distortion, while the vacancy alters the local environment of the two neighboring Ti from octahedral to pyramidal, and leads to a visible relaxations of the eight closest oxygens. 

\begin{figure}[ht]
\begin{center}
\includegraphics[angle=0,width=9cm]{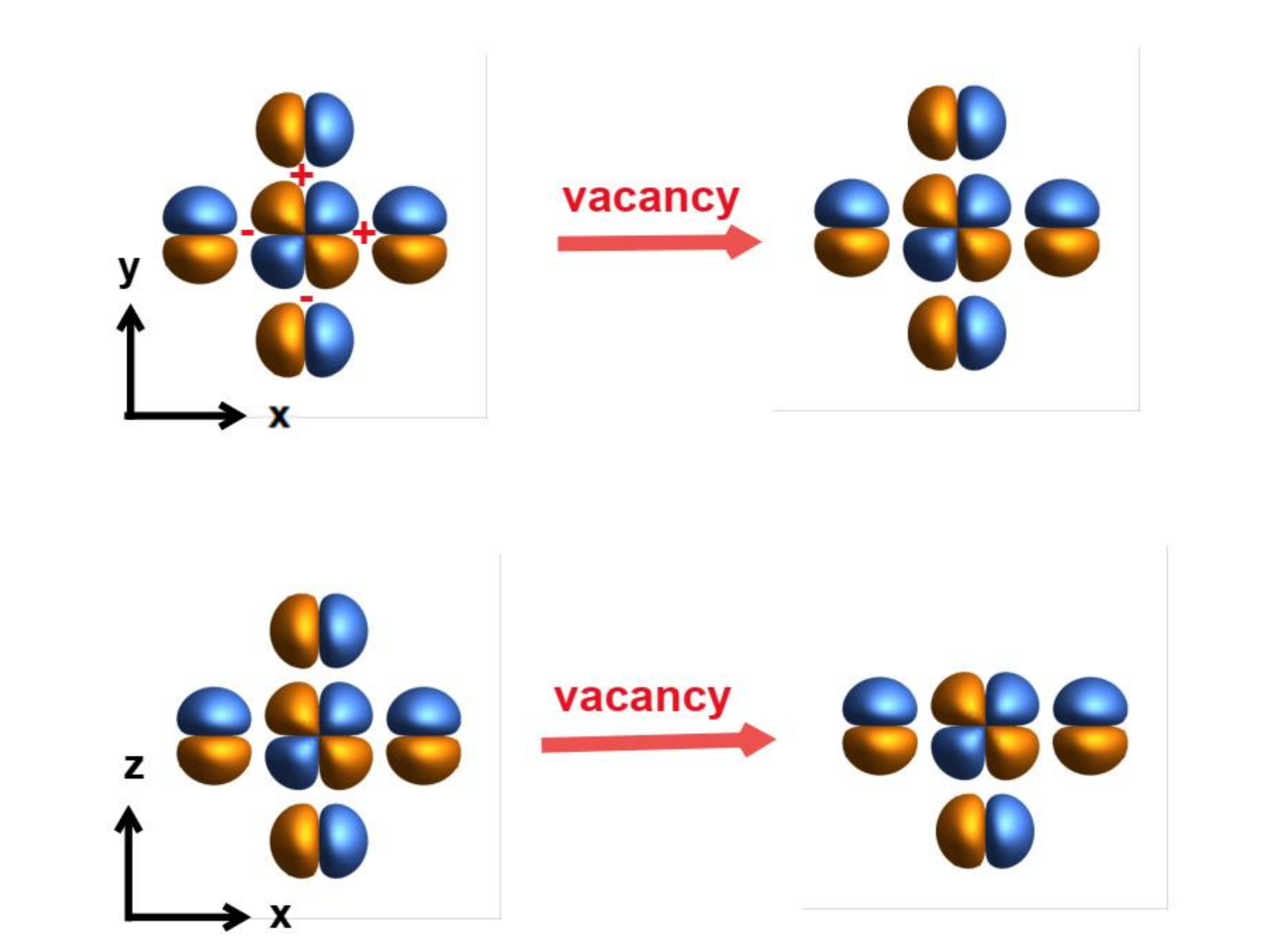}
\caption{\textbf{Effect of a tetragonal'' v-O vacancy on the ligand crystal field:} 
Relevant Ti $d$ and O $p$ wave functions for the wave vector {\bf k} corresponding to the $\Gamma$ point, in the real space around a Ti site. In absence of O vacancy, the hybridization of all $t_{2g}$ orbitals with the O states cancels out by symmetry. Introducing a vacancy does not change this for the $xy$ orbital (top) but the cancellation is broken by the vacancy for the $xz$ and $yz$ (not shown) orbitals (bottom). As a consequence, in the momentum space, the corresponding bands at the $\Gamma$ point shift upward.}
\label{xyxz}
\end{center}
\end{figure}

Neglecting the small distortion of the octahedra, at the $\Gamma$ point, which is the only one of interest, the hybridization of a Ti $t_{2g}$ orbital with the four neighboring O atoms cancels out by symmetry (Fig. \ref{xyxz}). However, if one apical oxygen is removed, the ligand field of the opposite one is uncompensated, so the corresponding band state is pushed up. In our case, this means that the $xz$ and $yz$  bands will be pushed up closer to the (unaffected) $xy$ band {\it at the $\Gamma$ point} (but not at the Z point). Of course, given the delocalization of the doped electrons, the effect on the entire band is reduced compared to the actual change of the site-local crystal field, but it still rather large on the scale of interest (i.e., $\sim 10$ meV). Note that no extra electron density is found in the calculations near the vacancy site  (Fig. \ref{xyxz}).

The situation for a v-O vacancy is more complicated. Qualitatively, the perturbation of the Ti environment is the same as depicted in Fig. \ref{xyxz}b, but the picture needs to be rotated by 90$^\circ$. Now it is not the $xy$ orbitals that is being push down with respect to the other two, but $yz$. Since the $xz/yz$ degeneracy with $xy$ is already broken due to octahedra rotation, it leads to an interesting situation when the states at $\Gamma$ are threefold split already on the nonrelativistic level, $i.e.$, without spin-orbit.

%We have also performed calculations including spin-orbit, which we do not discuss here, because the results are extremely sensitive to the actual separation between the $t_{2g}$ level, which as mentioned above, is expected to be very concentration dependent for O vacancies, and different in the experimentally relevant concentration range from the one obtained in these $8\times 8\times 8$ calculations.

Interestingly, the calculated threshold for the first Lifshitz transition in the pristine crystal  \cite{marel2011} is n$_{c1}=6.4 \times 10^{17} cm^{-3} $. This is in quantitative agreement with what we find experimentally for SrTi$_{1-x}$Nb$_x$O$_{3}$ (n$_{c1}= 6.5 \pm2 \times 10^{17} cm^{-3} $; Fig.\ref{FigQO2}a), confirming that the rigid band approximation holds in this case (but not in SrTiO$_{3-\delta}$).

 \section{Discussion}
 
Pristine strontium titanate goes through a cubic-to-tetragonal structural phase transition \cite{Lytle} at $T_s \simeq 105$ K. This is a (remarkably) second-order \cite{Salje_1998} structural phase transition driven by the softening of a Transverse Acoustic (TA) mode at the zone boundary \cite{Shirane1969}. Previous experiments have found that $T_s$ increases by Nb substitution and decreases by oxygen deficiency \cite{BAUERLE19781343,Tao2016}. In oxygen-reduced samples with a carrier density of $0.5\%$ per formula unit, $T_s$ \textit{drops by} $\sim 10$K \cite{BAUERLE19781343}. In Nb-doped samples with a carrier density of $1\%$ per formula unit, $T_s$ \textit{increases by }$\sim 20$K \cite{Tao2016}. This implies that the effect of the two doping routes on the boundary TA mode. This trend is reproduced by our DFT calculations. They found, as reported in the previous section, that the energy difference between the cubic and the tetragonal structures is amplified by Nb substitution and diminished by oxygen removal. Given the link between $T_s$ and energy difference, our calculations provide a qualitative account of the experimental observation.
 
The softening of another mode, a Transverse Optical (TO) phonon located at the center of the Brillouin zone, would lead to ferroelectricity. This mode is suspected to play a prominent role in driving the  superconducting instability in the dilute limit \cite{Marel2019,Kiselov2021,Volkov2022,Gastiasoro2022,Yu2022}. Given the contrasting consequences  of the two doping routes for the boundary TA phonon (see above), one may wonder what would be their effect on the zone-centered ones. The dispersion of these TO phonons has been found to evolve with doping \cite{Collignon2020,Rehwald1970}. But there is no record of a comparative study of their evolution in oxygen-reduced and Nb-substituted samples. Future studies may discriminate between the fingerprints of the two doping routes. 

The present study reveals the contrasting outcomes of Nb substitution and oxygen reduction for metallicity and superconductivity. As we saw, the experimental data reveals a difference in the Fermi surface geometry in the two cases near the first Lifshitz transition. This is in qualitative agreement with our DFT calculations. They show that the band dispersion is different in the two cases. This agreement constitutes a `proof of the concept'. A quantitative confrontation between theory and experiment is beyond what can be achieved by the state-of-the-art. Indeed, the theoretical supercell is equivalent to a doping concentration of $\sim 10^{-2}$ per formula unit. This exceeds by two orders of magnitude the experimental one and impedes a direct comparison.

Let us note that very recently, Tomioka \textit{et al.} \cite{Tomioka2022} have confirmed that below a threshold  density of  $\approx 8 \times 10^{18}$ cm$^{-3}$,  SrTi$_{1-x}$Nb$_x$O$_{3}$ and Sr$_{1-x}$La$_xTi$O$_{3}$ do not superconduct. On the other hand, replacing Sr with either Ca or Ba makes them superconducting. This provides additional support for dopant dependent superconductivity in the dilute limit and calls for an experimental  study of the fermiology in polar metals such as Nb:Sr$_{0.985}$Ca$_{0.015}$TiO$_3$.

Dopant-dependent superconductivity has been reported in other cases. Beyond a critical threshold of Tl doping, Pb$_{1-x}$Tl$_x$Te is a superconductor, but Pb$_{1-x}$Na$_x$Te is non-superconducting \cite{Matsushita2005,Giraldo2018}. The difference has been tracked to the presence of Tl impurity states near the Fermi energy. Let us note that the concentration of dopants is two orders of magnitude lower in strontium titanate than in lead telluride. The experimental observation of dopant dependent superconductivity has motivated theoretical scenarios \cite{Dzero2005,Kivelson2020}, which may be relevant to our result. Nevertheless,  large Hubbard correlations are unlikely in the extreme dilute limit. Occam's razor \cite{Mazin2022} would favor scenarios which do not require them. 

Thus, like lead telluride, strontium titanate demonstrate that the rigid band approximation does not capture the whole physics when a small concentration of dopants turns a solid off-stoichiometric. Future studies will tell if the breakdown of the rigid band approximation has any visible signature in the thermoelectric response \cite{Lee-2012}, as found in the case of PbTe \cite{Heremans_2008}, where doping by Tl has led to an amplification of the thermoelectric figure of merit.
\begin{table*}[ht]
    \centering
    \begin{tabular}{|c|c|c|c|c|c|c|c|}
    \hline
      Sample & $\mu_H$  & n$_{H}$  & F$_1$ & F$_2$ &  target used & thickness & Undoped cap layer \\
       & V.cm$^{-2}$.s$^{-1}$ & $cm^{-3}$ & T & T &   & nm& nm \\
      \hline
      $S_{1}$ &  4100 & 4.4e18 & 36 & 137  & 0.04 at.\% Nb & 892&  119\\
      $S_{2}$  &  3162 &  1.65e18 & 16 & 72  &  0.02 at.\% Nb& 617&  102\\
      $S_{3}$  &  1365 & 1.5e18 & 14 & 58 &  0.02 at.\% Nb& 895&  90\\
     $S_{4}$  &  9084 & 8.6e17 & 9 & 33 &  0.02 at.\% Nb +undoped & 748&  100\\
        \hline
    \end{tabular}
    \caption{\textbf{SrTi$_{1-x}$Nb$_x$O$_{3}$ thin films studied in this work and their properties-} Carrier densities and Hall mobilities are given at 2 K.}
    \label{tab:Sample}
\end{table*}

In summary, we found that doping strontium titanate by removing oxygen leads to a dilute metal different from the one obtained by substituting Ti with Nb. The two detected differences are the band dispersion and the presence and absence of a superconducting instability. The exceptionally dilute superconductivity  is an instability in a band specifically sculpted by oxygen vacancies of strontium titanate.

 \section{Materials and methods}
Nb-doped films were synthesized on SrTiO$_3$ (001) substrate in pulsed laser deposition, as detailed previously \cite{Inoue2019,Kozuka2010}. The films, which are listed in table \ref{tab:Sample}, are all sufficiently thick to host a three-dimensional Fermi surface as found in a previous study of their fermiology \cite{Kim2011}. 

Oxygen-deficient SrTiO$_3$ single crystals were obtained by annealing commercially bought substrates in a manner similar to what was described in ref.\cite{Spinelli2010,Lin2013,Lin2014}. Four-contact resistivity measurements were performed in a dilution refrigerator inserted in a 17 T superconducting magnet. 

Band calculations used the Projector Augmented Wave method together with the Generalized Gradient Correction to the exchange-correlation potential \cite{PBE}, as implemented in the VASP package \cite{vasp}. Fig. 5 was generated using the VESTA program \cite{vesta}.

 \section{Acknowledgements}
 We thank Claude Ederer and Maria Gastiasoro for useful discussions. 
 
This work was supported by the Agence Nationale de la Recherche (ANR-18-CE92-0020-01; ANR-19-CE30-0014-04), by Jeunes Equipes de l$'$Institut de Physique du Coll\`ege de France and by a grant attributed by the Ile de France regional council. It was funded in part by a QuantEmX grant from ICAM and the Gordon and Betty Moore Foundation through Grant GBMF9616 to KB. The work at SLAC/Stanford is supported by the U.S. Department of Energy, Office of Basic Energy Sciences, Division of Materials Sciences and Engineering, under Contract No. DE-AC02-76SF00515. 
\bibliography{STODopant}

\end{document}